\documentclass[trackchanges,twocolumn]{aastex701}

\usepackage{amsmath}
\usepackage{xcolor}
\usepackage{mathrsfs}
\usepackage{multirow}
\usepackage{makecell}
\usepackage{placeins}
\usepackage{gensymb}

\shorttitle{Schmidt Observatories Multimessenger prospects}
\shortauthors{Kunnumkai et al.}
\received{\today}
\revised{}
\accepted{}
\submitjournal{ApJ}
\graphicspath{{./}{Figures/}}

\newcommand{\mcwilliams}{
    McWilliams Center for Cosmology and Astrophysics,
    Department of Physics,
    Carnegie Mellon University,
    5000 Forbes Avenue, Pittsburgh, PA 15213, USA
}

\begin{document}

\title{Characterizing Gravitational Wave Electromagnetic Counterparts with the Schmidt Observatory System}
\author[0009-0000-4830-1484]{Keerthi Kunnumkai}
\affiliation{\mcwilliams}
\email[show]{kkunnumk@andrew.cmu.edu}  

\author[0000-0002-6011-0530]{Antonella Palmese}
\affiliation{\mcwilliams}
\email{apalmese@andrew.cmu.edu}

\author[0000-0002-4043-9400]{Thomas Wevers}
\affiliation{Astrophysics \& Space Center, Schmidt Sciences, New York, NY 10011, USA}
\email[]{twevers@schmidtsciences.org} 

\author[0000-0002-8255-5127]{Mattia Bulla}
\affiliation{Department of Physics and Earth Science, University of Ferrara, via Saragat 1, I-44122 Ferrara, Italy}
\affiliation{INFN, Sezione di Ferrara, via Saragat 1, I-44122 Ferrara, Italy}
\affiliation{INAF, Osservatorio Astronomico d’Abruzzo, via Mentore Maggini snc, 64100 Teramo, Italy}
\email[]{mattia.bulla@unife.it}

\author[0000-0003-3703-5154]{Suvi Gezari}
\affiliation{Department of Astronomy, University of Maryland, College Park, MD, 20742-2421, USA}
\email[]{suvi@umd.edu}

\author[0000-0002-4436-4661]{Saul Perlmutter}
\affiliation{Physics Division, E.O. Lawrence Berkeley National Laboratory, 1 Cyclotron Rd., Berkeley, CA, 94720, USA}
\affiliation{Department of Physics, University of California Berkeley, Berkeley, CA 94720, USA}
\email[]{saul@lbl.gov} 
 
\author[0000-0002-9396-7215]{Liliana Rivera Sandoval}
\affiliation{Department of Physics and Astronomy, University of Texas Rio Grande Valley, Brownsville, TX 78520, USA}
\affiliation{South Texas Space Science Institute, University of Texas Rio Grande Valley, Brownsville, TX 78520, USA}
\email[]{liliana.riverasandoval@utrgv.edu}

\author[0000-0001-7409-5688]{Gudmundur Stefansson}
\affiliation{Astrophysics \& Space Center, Schmidt Sciences, New York, NY 10011, USA}
\affiliation{Anton Pannekoek Institute for Astronomy, 904 Science Park, University of Amsterdam, Amsterdam, 1098 XH}
\email[]{gstefansson@schmidtsciences.org}

\author[0000-0002-8121-2560]{Mickael Rigault}
\affiliation{Universit\'e Lyon 1, CNRS, IP2I Lyon, UMR 5822, Villeurbanne, France}
\email[]{m.rigault@ip2i.in2p3.fr} 

\author[0000-0002-5463-9980]{Arpita Roy}
\affiliation{Astrophysics \& Space Center, Schmidt Sciences, New York, NY 10011, USA}
\email[]{aroy@schmidtsciences.org} 

\begin{abstract}
The emergence of gravitational wave (GW) multimessenger astronomy, marked by the detection of GW170817 and its associated kilonova, has opened a new window into the physics of compact binary mergers. Building a statistical population of kilonovae remains a challenge, limited by the low neutron star merger detection rates and the rapid evolution of these faint transients. The former problem will be overcome with upcoming LIGO/Virgo/KAGRA (LVK) sensitivity improvements leading to the fifth Observing Run (O5), while the latter will be addressed by the Schmidt Observatory System (including the Argus Array, the Deep Synoptic Array (DSA), and the Lazuli Space Observatory), in addition to ongoing and upcoming surveys (including the Rubin Observatory Legacy Survey of Space and Time). Using simulations of expected kilonovae at LVK O5 sensitivity, we find that Argus and Rubin could detect binary neutron star kilonovae at a rate of 0.3-4.2 yr$^{-1}$ and 0.6-7.0 yr$^{-1}$ respectively. For neutron star-black hole mergers, Rubin could detect up to 1 kilonova per year. Lazuli will be able to characterize this kilonova population down to $g \sim 26$ mag as early as 12 hours post merger. Additionally, $\sim35\%$ of these events are expected to exhibit a jet afterglow that will be detectable by DSA. We further characterize kilonova spectra to determine observable features of heavy element production. Finally, we develop an observing strategy that combines high cadence imaging and spectral monitoring, maximizing Lazuli's ability to constrain ejecta properties and composition of GW counterparts with reasonable observing time.
\end{abstract}


\section{Introduction} 

The detection of GW170817 by the LIGO and Virgo collaborations marks the first gravitational wave (GW) observation of a  binary neutron star (BNS) merger \citep{2017Abbottmma,GW170817abbott}. This event was followed by GRB~170817A $\sim 1.7$ s after the merger \citep{Goldstein_2017,Savchenko_2017}, by kilonova emission after $\sim 11$ hours \citep{Arcavi2017,Chornock2017,Coulter2017,Covino2017,Cowperthwaite2017,Drout2017,Evans2017, Kasliwal2017, Lipunov2017,Nicholl2017,Pian2017,Shappee2017,Smartt2017,Santos2017,Tanvir2017,Utsumi2017,Valenti2017}, and by radio and X-ray afterglow emission on timescales of days to weeks \citep{Margutti2017,Troja2017,Hallinan2017,Alexander2017,Haggard2017,Lamb2017jet,Lazzati2018,Resmi2018,Mooley2018}. Kilonovae trace the neutron-rich ejecta from the merger, while the afterglow traces the relativistic outflow launched from the remnant. Observing these electromagnetic (EM) counterparts allows us to probe heavy element production \citep{2017Kasen,Pian2017}, jet physics \citep{Troja2017,Mooley2018,Alexander2018,Ghirlanda2019,Troja2020,Ryan2024}, physics of neutron star mergers \citep{Coughlin_2018, Margalit_2019, Dietrich_2020} and cosmology \citep{gw170817_nature,Hotokezaka19,2022Bulla,Palmese:2023beh}, as well as the nature of the progenitor and remnant object \citep{Yu_2018}.

Neutron star mergers have long been proposed as promising sites for the synthesis of rapid neutron capture elements via r-process nucleosynthesis \citep{1974Lattimer,1989Eichler,2017Metzger}. Characterizing the composition of the kilonova ejecta that follows such a merger provides a unique observational window into the process. While photometry can constrain the broad luminosity and the color evolution of a kilonova, which can inform our understanding of its ejecta mass, velocity and electron fraction, spectroscopy provides a more direct way to probe the ejecta composition. Correctly identifying kilonova spectral lines remains a challenge because rapidly expanding ejecta possess a velocity gradient, which Doppler shifts the individual atomic transitions by different amount in different layers of the ejecta. This will broaden the line and cause lines with similar wavelengths to overlap and blend \citep{2017Kasen,2020Tanaka,2021Gillanders,2022Gillanders}. As a result, the same spectral features can support competing identifications. In AT2017gfo, for example, \citet{Smartt2017} attributed the early absorption feature {at 1.8 $\mu$m} to Te\,\textsc{I} and Cs\,\textsc{I}, both second r-process peak elements, while \citet{2019Watson} argued that Sr\,\textsc{ii}, a first r-process peak element, explains the same feature better.

Identifying the time at which a feature appears and tracing its subsequent evolution can help reduce ambiguities among blended features. This requires spectroscopic sequences that span optical to infrared wavelengths, since the kilonova cools and its emission shifts redward over time \citep{2022Gillanders, 2024Sneppen}. AT2017gfo provided such a sequence because its proximity ($\sim 40$ Mpc) made repeated high quality spectroscopy possible as the transient rapidly faded \citep{2017McCully, Pian2017, Smartt2017}. Moreover, the optical counterpart was rapidly identified (within hours of the GW trigger). Obtaining comparable observations for future events will be difficult as the increasing sensitivity of LIGO/Virgo/KAGRA (LVK) GW detectors will extend neutron star merger detections to distances of hundreds of Mpc, where their kilonova will appear much fainter. Additionally, GW localization regions can span tens to thousands of deg$^2$, and a counterpart must first be identified through sensitive wide field imaging before spectroscopy can begin, so this may in general take longer  than a few hours. Thus, obtaining spectroscopic sequences will become challenging for most ground based instruments, even at peak luminosity.

Beyond thermal emission, afterglow emission is also produced when a successfully launched jet interacts with the surrounding medium and generates synchrotron emission. If the jet is viewed close to its axis, the prompt gamma ray burst (GRB) and early afterglow are dominated by the jet core, making early optical and near-infrared monitoring important on timescales of hours to about a day after the merger. Off-axis events, however, are expected to constitute the majority of detectable GW sources, given that GW emission is sensitive to the inclination angle of the binary with respect to the observer. As a result, the viewing angle of the detectable GW source population peaks close to $\sim 30$ deg \citep{Schutz2011}, resulting in a slowly-evolving jet afterglow peaking in luminosity hundreds of days after merger \citep{2026Kaur}, which is close to what was observed for GW170817 \citep{Margutti_2021,
Palmese:2023beh}.

The Eric and Wendy Schmidt Observatory System\footnote{\url{https://www.schmidtsciences.org/schmidt-observatory-system/}} is expected to provide new opportunities within this landscape of multi-messenger discovery and characterization in the coming years \citep{RoyLazuli2026, 2026Wevers}. By combining the large aperture equivalent, wide-field optical survey to be carried out by the Argus Array\footnote{\url{https://argus.unc.edu}} \citep{argus}, with radio observations from the Deep Synoptic Array\footnote{\url{https://www.deepsynoptic.org}} (DSA; \citealt{hallinan2019dsa2000radiosurvey}), rapid spectroscopic observations from the Large Fiber Array Spectroscopic Telescope\footnote{\url{https://astro.arizona.edu/research-groups/lfast}} (LFAST), as well as spectroscopic and imaging space-based data from the Lazuli Space Observatory \citep{RoyLazuli2026}, the System provides a unique opportunity for coordinated discovery and characterization of multimessenger sources. In addition, the NSF-DOE Vera C. Rubin Observatory Legacy Survey of Space and Time (LSST; \citealt{2019ApJ...873..111I}) has recently started and is expected to discover EM candidate counterparts to GW events that can be further observed and characterized with this Observatory System.

In this work, we simulate BNS and neutron star-black hole (NSBH) merger GW events as expected during the LVK O5 observing run and as informed by available GW population studies. The O5 run is expected to start in $\sim 2029$ and last about 3 years\footnote{\url{https://emfollow.docs.ligo.org/userguide/capabilities.html}}, and will have significant overlap with the Schmidt Observatory System expected operations. We attach kilonova and afterglow emission to the simulated GW events and evaluate their detectability with Argus and Rubin. For the detectable population, we estimate the Lazuli exposure times (ETs) needed for follow-up at different epochs. We then use AT2017gfo-like spectral models to determine when key spectral features are expected to be observable within Lazuli's wavelength range, with a goal of developing possible observing strategies for future BNS and NSBH detections. Additionally, we estimate the potential for jet afterglow detections, including with DSA. 

We assume that the counterpart has been identified throughout this work; i.e. we do not simulate the process of identifying the right counterpart in a stream of transient alerts.

The manuscript is structured as follows. In Section \ref{sec:observatories} we introduce the Observatories considered in this work. In Section \ref{sec:sims} we describe how the GW and kilonova simulations are produced. In Section \ref{sec:results} we show detectability rates for kilonovae and afterglows across observatories, and present prospects for photometric and spectroscopic campaigns aimed at the characterization of EM counterparts with Lazuli. Finally, in Section \ref{sec:conclusions} we outline our conclusions.

\section{Observatories}\label{sec:observatories}

Four main projects are being developed as part of the Schmidt Observatory System. The Argus Array \citep{argus}, planned to start in 2028, is a ground-based optical observatory that combines 1,212 small aperture telescopes into a single 8-meter equivalent telescope, yielding an instantaneous field of view of 8,000 deg$^2$ and a nightly coverage of 21000 deg$^{2}$ over the Northern sky in $\beta$ ($g'$-like) and $\rho$ ($r'$-like) bands\footnote{https://github.com/argus-hdps/argus$\_$sim/} (Corbett et al., in prep.). The expected average depths from 1 hour, 1 night, 1 week, and 6 months co-adds are 22.9, 23.9, 24.7, and 26.5 mag, respectively\footnote{https://simulators.schmidtobservatorysystem.org/argus/}. In what follows, we consider the 1-night co-add depth for kilonova identification, as this maximizes our ability to detect both faint and fading sources. Longer co-adds are useful for the afterglow component from off-axis mergers.

The Argus Array survey will be complementary to the Rubin LSST, which is being carried out from Chile and is expected to continue its main survey until 2036. Rubin provides the greater depth needed to recover fainter and more distant counterparts across the typical LVK localization regions. Rubin observes in $ugriz$ bands and will dedicate a few per cent of its survey time to target of opportunity (ToO) observations, including GW events \citep{2024AndreoniRubin}. The Rubin ToO strategy for BNS and NSBH mergers is motivated by the need to cover $100-500~ \rm deg^{2}$ skymaps while reaching kilonova relevant depths at $300 - 700$ Mpc. For 30 s exposures, the expected Rubin depths are $g \sim 25.0$, $r \sim 24.7$ and $i \sim 24.0$, with deeper limits possible for 120 s ($g \sim 25.8$, $r \sim 25.5$ and $i \sim 24.8$) and 180 s exposures ($g \sim 26.0$, $r \sim 25.7$ and $i \sim 25.0$) (see Figure \ref{fig:lcbns_nsbh_r_g}). Unless otherwise specified, we use the 30 s ET while considering if a kilonovae or afterglow is detectable by Rubin. For well localized events ($\Omega < 500 ~\rm deg^2$), \citet{2024AndreoniRubin} proposes a high-cadence observing strategy where the triggering is only recommended when probability of an alert being a BNS+NSBH $\ge$ 90\%. Argus and Rubin combined will provide access to the full observable extragalactic sky, maximizing our ability to identify early optical counterparts to GW events, which can then be further characterized with other instruments. 

The Deep Synoptic Array \citep{hallinan2019dsa2000radiosurvey} is a Schmidt Observatory System radio array that will start operations in 2029 from Nevada (USA). Survey operations will repeatedly cover 31,000 deg$^2$, serendipitously enabling the identification of jet afterglow radio emission from GW sources, in addition to pointed observations of EM counterparts identified by other surveys (including Rubin LSST and Argus). DSA will observe at 0.7-2 GHz (we consider 1.4 GHz as a fiducial frequency in what follows) with a typical continuum sensitivity rms of 0.6 $\times 10^{-3}$ mJy with a 1 hour integration time\footnote{https://www.deepsynoptic.org/sensitivity/index.html}.

The Schmidt Observatory LFAST \citep{berkson_large_2024} will deliver medium to high resolution optical to near-infrared (NIR) fiber-fed spectroscopy. The prototype, which will operate over a timescale relevant for this work, will comprise of 20 modules which will be incoherently combined, achieving an equivalent collecting area of 3.5 m$^2$. We do not consider this specific observatory in more detail in this study, since we focus on spectroscopic observations of transients that are typically too faint for the prototype.

Finally, the Lazuli Space Observatory \citep{RoyLazuli2026} is also part of the Observatory System, and combines a 3m space-based telescope with an integral field spectrograph (IFS) for continuous 0.4-1.7 $\mu$m spectroscopy at $R \sim 100-500$. The IFS uses an image slicer to divide a small field on the sky into slices, reformat them into a pseudo-slit and disperses the light with a prism onto a near infrared detector, allowing each observation to capture the optical and near-infrared spectrum at the same time \citep{RoyLazuli2026}. Lazuli also includes a Widefield Context Camera (WCC) for diffraction limited multiband imaging from 0.35-1.0 $\mu$m over a $35^\prime \times 12^\prime$ footprint; the focal plane is tiled with 23 CMOS detectors (IMX455 sensors with an on-sky area of 2.7$\times$1.8 arcmin and HWK4123 sensors covering 1.5$\times$0.9 arcmin. The WCC makes use of broadband filters $ugriz$, as well as broad filters covering most of the 0.4-0.9 $\mu$m range and narrow-band filters \citep{RoyLazuli2026}. For kilonova and afterglow studies, WCC monitoring can help separate thermal kilonova emission from non-thermal synchrotron afterglow emission \citep{2026Wevers}.

Unlike wide-field surveys such as Argus and Rubin, which are optimized for discovery, Lazuli is designed for rapid space-based characterization once a compelling transient is identified \citep{2026Wevers}. Compared to ground-based follow up, Lazuli avoids atmospheric absorption and weather losses and is designed around a rapid ToO response requirement of $<4$ hours \citep{RoyLazuli2026}. 

\section{Simulations}\label{sec:sims}
\subsection{Gravitational wave simulations}
\label{sec:gw_sim}

We simulate gravitational wave events using the framework described in  \citet{Kunnumkai_sim_2025}. We use tools from \texttt{LALSuite} \citep{lalsuite,swiglal} to draw GW injections according to the specified population model and compute the detector response for the assumed network configuration. Unlike \citet{Kunnumkai_sim_2025}, which drew source parameters from \textsc{PowerLaw+Dip+Break} (PDB) \citep{2022Farah,2020Fishbach} analysis based on the GWTC-3 catalog \citep{2023Abbottgwtc3cat}, we adopt the maximum a posteriori (MAP) hyperparameter values of the \textsc{FullPop-4.0} distribution based on the GWTC-4 catalog \citep{LVK2025gwtc4pop}. The \textsc{FullPop-4.0} model is based on the \textsc{PowerLaw+Dip} \citep{2020Fishbach}, \textsc{BrokenPowerLaw+Dip} \citep{2022Farah} and \textsc{MultiPDB} \citep{2025MaliEssick} frameworks. Unlike the PDB model, \textsc{FullPop-4.0} allows an upper mass gap, and employs separate pairing functions for NSs containing binaries and for binary black holes (BBHs). The pairing function transitions at $m_2 = 5~M_{\odot}$, allowing distinct mass-ratio behavior for NS-bearing mergers and BBHs. The \textsc{FullPop-4.0} model represents the low mass black hole (BH) peak near $9~M_{\odot}$ with a Gaussian component whose lower mass tail extends to the $3-5~M_{\odot}$ region. Together with the underlying mass distribution, this tail permits a non-zero probability of having objects in this region, removing the need to impose a separate mass gap between NSs and BHs. The component masses $(m_1,m_2)$ are drawn from this MAP population model, and we adopt the convention $m_{1} > m_{2}$. 

Sources are distributed uniformly in comoving volume, while their sky positions and binary orientations are sampled isotropically. Waveforms are generated using tools from \texttt{LALSuite} \citep{lalsuite} and projected onto the detector network assuming the upcoming LIGO/Virgo/KAGRA (LVK) O5 run detector sensitivities. For LIGO detectors, we assume A+ sensitivity, avirgo\_O5high\_NEW.txt for Virgo, and kagra\_80Mpc.txt for KAGRA\footnote{all sensitivity curves are taken from \url{https://dcc.ligo.org/LIGO-T2000012-v1/public}}. A duty cycle of 70\% is assumed for each detector. We consider an event to be detected if it satisfies:
\begin{equation}
\rho_{\rm net} \ge 8 \quad \text{and} \quad \rho_{\rm det} \ge 1,
\end{equation}
where $\rho_{\rm net}$ is the network signal to noise ratio (SNR) and $\rho_{\rm det}$ is the SNR in at least 1 detector \citep{Petrov_2022,Kunnumkai_sim_2025}. For all detected events, we compute rapid sky localizations using \texttt{BAYESTAR} \citep{bayestar,Singer}, which provides the localizations to be used to assess the feasibility of EM counterpart follow-ups.

Since the FullPop-4.0 model does not explicitly assign astrophysical classes, we delineate between NSs and BHs based on the Tolman-Oppenheimer-Volkoff (TOV) limit and rotational corrections described in \citet{Breu_2016,2020Most}. Under this prescription, a BNS is defined as having $m_1, m_2 < M_{\rm max}$ and an NSBH is defined as having $m_2 < M_{\rm max}$ and $m_1 > M_{\rm max}$, where $M_{\rm max}$ is the maximum mass a rotating NS can sustain. Throughout the study, we adopt the maximum \texttt{a posteriori} equation of state (EoS) from \citep{Huth:2021bsp}, which supports a TOV mass is 2.436 M$M_{\odot}$. 

To account for the absence of BNS detection in O4, we rescale the GW merger rate density described in \citet{LVK2025gwtc4pop}, i.e. 
\begin{align}
R_{\rm BNS} &= 89^{+159}_{-67} \ \mathrm{Gpc^{-3}\,yr^{-1}},
\label{eq:2}\\
R_{\rm NSBH} &= 23^{+20}_{-13} \ \mathrm{Gpc^{-3}\,yr^{-1}},
\end{align}

by multiplying the BNS merger rate by B = 0.4396, derived using the BNS ranges and duty cycles described in \url{https://gwosc.org/detector_status/}. Here B is defined as:
\begin{equation}
    B = \frac{\sum{VT,_{\text{all observing runs}}}}{\sum VT, _{O1-O4a}}
\end{equation}
{where VT is used to denote the sensitive space time volume.} This calculation assumes that the BNS merger rate density does not evolve with redshift, and that the relative changes in VT are constant in mass.

\subsection{Kilonova simulations}
\label{kn_sim}

To generate the kilonova lightcurves for the simulated BNS and NSBH mergers, {we use kilonova models produced by \texttt{POSSIS} radiative transfer code \citep{Bulla_2019,Bulla_2023} and implemented into the Nuclear Multi-Messenger Astronomy (NMMA) framework \citep{Dietrich_2020,Pang_2023}.} The kilonova ejecta parameters, namely, those describing the dynamical ejecta (launched tidally before and during the merger) and the post merger wind ejecta (launched from the remnant accretion disk) are computed from a set of priors and binary merger parameters as described in \citet{Kunnumkai_sim_2025}. 

For BNS models, we use \texttt{Bulla2025MLP} \citep{Bulla_2023,2025Koehn}, which is a multilayer perceptron surrogate model trained on the latest \texttt{POSSIS} grid (see Section \ref{sec:kn_grid}). The \texttt{POSSIS} (and consequently, surrogate models) inputs include {dynamical and wind ejecta masses M$_{\rm dyn}$ and M$_{\rm wind}$, velocities v$_{\rm dyn}$ and v$_{\rm wind}$, electron fractions Y$_{\rm e, \rm dyn}$ and Y$_{\rm e, \rm wind}$, and inclination $\iota$. v$_{\rm dyn}$, v$_{\rm wind}$, Y$_{\rm e, \rm dyn}$ and Y$_{\rm e, \rm wind}$ are uniformly sampled following Table~I of \citet{2025Koehn}.}  We compute ejecta masses, M$_{\rm dyn}$ and M$_{\rm wind}$ which are related to the binary properties using the fitting formulae described below, and we account for the luminosity distance and viewing angle based on our GW simulation. The dynamical ejecta mass follows \citet{Kr_ger_2020}, as:
\begin{equation}
\label{eq:mdyn_bns}
    \frac{M_{\rm dyn}^{\rm BNS}}{10^{-3}} = \left (\frac{a_{\rm dyn}}{C_1} + b_{\rm dyn} \left (\frac{m_2}{m_1} \right)^{n} +c_{\rm dyn}C_{1}\right)m_{1} + (1 \leftrightarrow 2).
\end{equation}
where $C_{1} = Gm_{1}/R_{1}c^{2}$ and $a_{\rm dyn}= -9.3335$, $b_{\rm dyn} = 114.17$, $c_{\rm dyn} = -337.56$ and $n = 1.5465$ are the best-fit parameters.

The BNS disk mass is computed using the fitting formula from \citealt{Dietrich_2020}:
\begin{equation}
    \label{eq:mdisk_bns}
    \log_{\rm 10}\left(M_{\rm disk}^{\rm BNS}\right) = \max \left (-3, a\left(1 + b \tanh \left [\frac{c-\frac{M_{\rm tot}}{M_{\rm th}}}{d} \right] \right) \right)\, .
\end{equation}
Here, $M_{\rm th} = k_{\rm th}M_{\rm TOV}$ \citep{Hotokezaka_2011} is the threshold for prompt collapse and $k_{\rm th}$ depends on the EoS and TOV mass. The mass ratio dependence is incorporated through $a = a_{0} + \delta_{a}x_{i}, b =  b_{0} + \delta_{b}x_{i}$ and $x_i = 0.5 \tanh(\beta(q - q_{t}))$. Here $a_{0} = -1.581$, $b_{0} = -0.538$, $c = 0.953$, $d = 0.0417$, $\delta_{a}=-2.439$, $\delta_{b} = -0.406$, $\beta=3.910$, and $q_{t}=0.900$. The wind ejecta mass $M_{\rm wind}^{\rm BNS}$ is $\zeta M_{\rm disk}^{\rm BNS}$ where $\zeta$ is the fraction of disk mass that contributes to the wind ejecta.

For NSBH models, we use \texttt{Bu2019nsbh} \citep{Bulla_2019, Anand_2020}. The dynamical ejecta is calculated from \citet{Kruger_2020} and follows the relation:
\begin{equation}
    \label{eq:M_dyn_nsbh}
    \frac{M_{\rm dyn}^{\rm NSBH}}{M_{\rm NS}^{b}} = a_{1}Q^{n_{1}}\frac{1-2C_{\rm NS}}{C_{\rm NS}} - a_{2}Q^{n_{2}}\frac{R_{\rm ISCO}}{m_1} + a_{4},
\end{equation}
where $a_1 = 0.007116$, $a_2 = 0.001436$, $a_4 =-0.02762$, $n_1 = 0.8636$, and $n_2 = 1.6840$ are the best fitting model parameters. $Q=m_1/m_2$ and $M_{\rm NS}^{b}$ = $m_2 \left(1 + \frac{0.6C_{\rm NS}}{1-0.5C_{\rm NS}}\right)$ \citep{Lattimer_2001} is the NS baryon mass. $R_{\rm ISCO}$ is the BH's inner most stable circular orbit radius. 

The NSBH remnant mass, in this case indicating the baryon mass outside the BH $\sim 10$ ms after the merger, is calculated following \citet{Foucart_2018} as:
\begin{equation}
    \label{eq:mdisk_nsbh}
    \hat{M}^{\rm NSBH}_{\rm rem} = \left [{\rm max}\left( a\frac{1-2C_{\rm NS}}{\eta^{1/3}} - b~R_{\rm ISCO}\frac{C_{\rm NS}}{\eta} + c ,0\right) \right]^{1+d}.
\end{equation}
Here, $\hat{M}$ = $M^{\rm NSBH}_{\rm rem}/M^{\rm b}_{\rm NS}$, $\eta$ = $Q/(1+Q^{2})$ refers to the symmetric mass ratio and $a=0.40642158$, $b=0.13885773$, $c=0.25512517$, $d=0.761250847$ are the best fitting parameters. Subtracting $M^{\rm NSBH}_{\rm dyn}$ from $\hat{M}^{\rm NSBH}_{\rm rem}$ gives the disk mass $M^{\rm NSBH}_{\rm disk}$, which when multiplied by $\zeta$ gives $M^{\rm NSBH}_{\rm wind}$.

After computing the ejecta parameters for each simulated BNS and NSBH event, the ones that produce sufficient ejecta (we assume {$10^{-5}~M_\odot$ for BNS and $10^{-4}~M_\odot$ for NSBH}) to produce a kilonova under our fiducial EoS are considered for further analysis. The resulting lightcurves computed from the kilonova surrogate model are compared with the adopted Argus and Rubin limiting magnitudes in each band to determine the fraction of kilonovae detectable by each survey.

\subsection{GW170817-like kilonova spectra}

The opacities taken from \citet{Tanaka2020} as input to \texttt{POSSIS} are a function of density, temperature, $Y_{e}$ and wavelength, but do not retain the atomic information needed to identify which elements contribute to the opacity in a given wavelength bin. Consequently, \texttt{POSSIS} spectra, which we use for estimating observing cadence and ET for following up Lazuli detectable kilonovae (see Section \ref{sec:kn_grid}) cannot directly support the identification of individual spectral features. To assess Lazuli's ability to recover kilonova spectral features, we instead adopt the AT2017gfo-like kilonova spectral models of \citet{2022Gillanders,2024Gillanders}, generated using \texttt{TARDIS} \citep{2014Kerzendorf,2022Gillanders}.
 
\texttt{TARDIS} is a Monte Carlo radiative transfer code that models emission from a spherically symmetric, homologously expanding ejecta above an optically thick inner boundary that represents the photosphere. The material above the photosphere constitutes the line forming region. Radiation emerging from the photosphere propagates through the expanding ejecta and interacts with the material through electron scattering and bound-bound and bound-free transitions. The emergent spectra are then built from the photon packets that escape the outer boundary. \citet{2022Gillanders} modeled AT2017gfo across 8 epochs from 0.5 days to 7.4 days postmerger. \texttt{TARDIS} is time independent and hence each epoch is modeled separately. The input parameters such as ejecta temperature, photospheric velocity, density profile and composition of the line forming region are evolved between epochs to reproduce the observed spectral evolution. 

In our analysis, we treat the \texttt{TARDIS} spectra as the intrinsic source frame spectra. For the luminosity distance under consideration, we transform the source spectra to the observer frame. The redshifted spectra for the respective epochs are then passed to \texttt{slicersim} \citep{rigault2026}, the spectral simulator and ET calculator of IFS. \texttt{slicersim} produces representative Lazuli observations of the input spectra at the adopted redshift and Lazuli wavelength range (0.4-1.7 $\mu$m) and determines the ET required at each epoch to achieve a specific SNR.

\subsection{Kilonova spectra for exposure time calculation}
\label{sec:kn_grid}

To estimate the exposure time Lazuli/IFS would require to follow-up kilonovae detectable by Argus and Rubin, \texttt{slicersim} takes the spectra for these events as input. For computational feasibility, events passing Argus and Rubin thresholds are associated with the spectrum of their nearest neighbor in the kilonova grid. 

To simulate the kilonova spectra, we use the latest version \citep{Bulla_2023} of the 3D Monte Carlo radiative transfer code \texttt{POSSIS} \citep{Bulla_2019}. Specifically, we use the most recent BNS and NSBH kilonova grids presented in \citet{2026Ahumada}, inspired from \citet{2023Anand} and \citet{2024Mathias}, respectively. This \texttt{POSSIS} version incorporates revised nuclear heating rates from \citet{Rosswog2024}, local time and density-dependent thermalization efficiencies and time- and wavelength-dependent opacities that depend on local temperature, density and composition \citep{Tanaka2020}. \texttt{POSSIS} propagates Monte Carlo photon packets through the homologously expanding ejecta, where the packets interact via bound-bound transitions and electron scattering.

\begin{table*}[!htpb]
\centering
\setlength{\tabcolsep}{4.5pt}
\renewcommand{\arraystretch}{1.85}
\begin{tabular}{c|c|c|cc|ccc|cc}
\hline
\multirow{3}{*}{Run} & \multirow{3}{*}{Binary type} &
\multirow{3}{*}{GW detection rate} &
\multicolumn{7}{c}{Kilonova detectability (yr$^{-1}$)} \\
\cline{4-10}
& & & \multicolumn{2}{c|}{Argus} & \multicolumn{3}{c}{Rubin} & \multicolumn{2}{c}{Argus+Rubin} \\
\cline{4-10}
& & (yr$^{-1}$) & $g$ & $r$ & $g$ & $r$ & $i$ & $g$ & $r$ \\
\hline

\multirow{2}{*}{O5} &
BNS  &
$4.95^{+7.38}_{-3.76}$ &
$1.37^{+2.04}_{-1.04}$ &
$1.68^{+2.50}_{-1.28}$ &
$2.37^{+3.53}_{-1.80}$ &
$2.64^{+3.94}_{-2.01}$ &
$2.34^{+3.49}_{-1.78}$ &
$2.87^{+4.29}_{-2.18}$ &
$3.27^{+4.88}_{-2.48}$ \\
\cline{2-10}
&
NSBH &
$13.56^{+11.66}_{-7.58}$ &
$0.09^{+0.08}_{-0.05}$ &
$0.11^{+0.09}_{-0.06}$ &
$0.34^{+0.29}_{-0.19}$ &
$0.40^{+0.34}_{-0.22}$ &
$0.33^{+0.28}_{-0.18}$ &
$0.35^{+0.30}_{-0.20}$ &
$0.41^{+0.35}_{-0.23}$ \\
\hline
\end{tabular}
\caption{O5 GW detection rates and kilonova detectability rates in each band for Argus and Rubin, accounting for no BNS detection in O4}
\label{tab:o5_rates_kn_detectability}
\end{table*}

The BNS kilonova grid is modeled with two axially-symmetric components: dynamical ejecta and wind ejecta. Following \citet{Bulla_2023} and \citet{2026Ahumada}, the dynamical ejecta density ($\rho$) and electron fraction ($Y_e$) assumed in our models depend on the polar angle $\theta$ as $\rho \propto \sin^{2} \theta$ and $Y_e \propto \cos^{2} \theta$, respectively. Such angular dependence concentrates neutron-rich, lanthanide-rich material towards the equatorial plane while higher $Y_{e}$, lanthanide-poor material dominates the poles \citep{2017perego,2018Radice,2023Setzer}. The wind ejecta are assumed to be spherically symmetric with a fixed electron fraction. The BNS kilonova grid is parameterized by six ejecta properties, {note that the BNS grid we describe here and in Section \ref{kn_sim} are the same}: the dynamical ejecta mass $M_{\rm dyn} = (0.001, 0.005, 0.01, 0.02)~M_\odot$, the dynamical mass-weighted average velocity $\bar{v}_{\rm dyn} = (0.12, 0.15, 0.20, 0.25)c$, the dynamical mass-weighted average electron fraction $\bar{Y}_{e,{\rm dyn}} = (0.15, 0.20, 0.25, 0.30)$, the wind ejecta mass $M_{\rm wind} = (0.01, 0.05, 0.09, 0.13)~M_\odot$, the wind mass-weighted average velocity $\bar{v}_{\rm wind} = (0.03, 0.05, 0.10, 0.15)c$, and the wind electron fraction $Y_{e,{\rm wind}} = (0.20, 0.30, 0.40)$. These parameter combinations result in 3072 ejecta models. Evaluated at 11 viewing angles $\theta_{\rm obs}$, uniformly spaced in $\cos\theta_{\rm obs}$ from face-on to edge-on orientations, the total number of BNS kilonovae realizations in the grid are 33,792. 

The NSBH ejecta are also modeled with dynamical and disk wind components. However, the angular distribution of the dynamical ejecta follows \citet{Kawaguchi_2020}, with a stronger concentration towards the merger plane, as described by $\rho \propto \frac{1}{1 + \exp[-20(\theta - 1.2)]}$. The composition in the grid is fixed, and set to $Y_e = 0.1$ for dynamical ejecta and $0.2 < Y_e < 0.3$ for wind ejecta \citep{Kawaguchi_2020, 2024Mathias}. The grid is constructed by varying binary properties and the neutron star EoSs. The BH mass $M_{\rm BH} = (4.0,  6.0,  8.0)~M_\odot$, BH spin $\chi_{\rm BH} = (0.0, 0.3, 0.6)$, NS mass $M_{\rm NS} = (1.2, 1.4, 1.6, 1.8)~M_\odot$ are the parameters that are varied within the grid, for the EoSs DD2 \citep{2010Hempel}, AP3 \citep{1998Akmal} and SFHO+H$\Delta$ \citep{2014Drago}. The properties of the ejecta, like mass and velocity, are obtained through prescriptions developed in \citet{2024Mathias}. Only 37 parameter combinations produce ejecta and therefore a kilonova. Evaluating these models at 11 viewing angles results in a total of 407 NSBH kilonova realizations. 

\subsection{Afterglow simulations}

To simulate afterglow emission from relativistic jets associated with BNS mergers, we use the publicly available \texttt{afterglowpy} package \citep{2020Ryan,2024Ryan} and follow the methodology of \citet{2026Kaur}. \texttt{afterglowpy} models GRB jet afterglow lightcurves by computing synchrotron radiation from a forward shock propagating into a uniform density interstellar medium. For each simulated BNS merger, we use the luminosity distance and binary inclination angle inferred from the GW simulation. Assuming that the jet axis is aligned with the binary orbital momentum, the inclination determines the jet viewing angle $\theta_{\rm v}$. We determine whether each merger successfully launches a relativistic jet based on the fate of the post-merger remnant, and we require a disk mass $M_{\rm BNS}^{\rm disk} > 10^{-3} M_{\odot}$ and a total binary mass $M_{\rm tot} > 1.2M_{\rm TOV}$. {For jet launching, we require that the post-merger remnant is either a hypermassive neutron star (a short-lived NS remnant that collapses to BH within tens of hundreds of milliseconds after the merger \citealt{Siegel_2013}) or promptly collapses to a BH. This assumption is motivated by numerical relativity simulations which shows that remnants that are long lived have the potential to drive sustained baryon loaded winds along the poles. This suppresses jet's Lorentz factor, which could prevent successful jet breakout on the short GRB timescales \citep{1999Ruffert,2002Rosswog,Dessart_2009,Murguia-Berthier_2014,Murguia-Berthier_2017, 2017Ciolfi}}. For each jet launching event, we perform Monte Carlo simulations by drawing N=100 samples from the distributions of afterglow parameters, including the electron fraction $Y_{e}$, isotropic equivalent core kinetic energy and viewing angle of the observer $\theta_{v}$. Afterglow lightcurves are then computed at each telescope's observing frequency between $10^{-2}$ days and 6.5 years post-merger in 100 logarithmically spaced time steps. The resulting flux at each time step is compared against the sensitivity limits of the observatories considered in this work (in addition to a few others) spanning radio through X-ray wavelengths, adopting the fiducial (``run 7'') set up from \citet{2026Kaur}. This set up assumes a Gaussian jet structure with a very narrow log-normal distribution for the jet core half opening angle, centered around 3.4 degrees. We adopt a Gaussian jet structure for the angular kinetic energy profile and an angular Lorentz factor profile with an initial core Lorentz factor $\tau_{0} = 300$.

\section{Results}\label{sec:results}
\subsection{Kilonova detectable sources}

\begin{figure*}[htpb!]
\centering
\includegraphics[scale = 0.56]{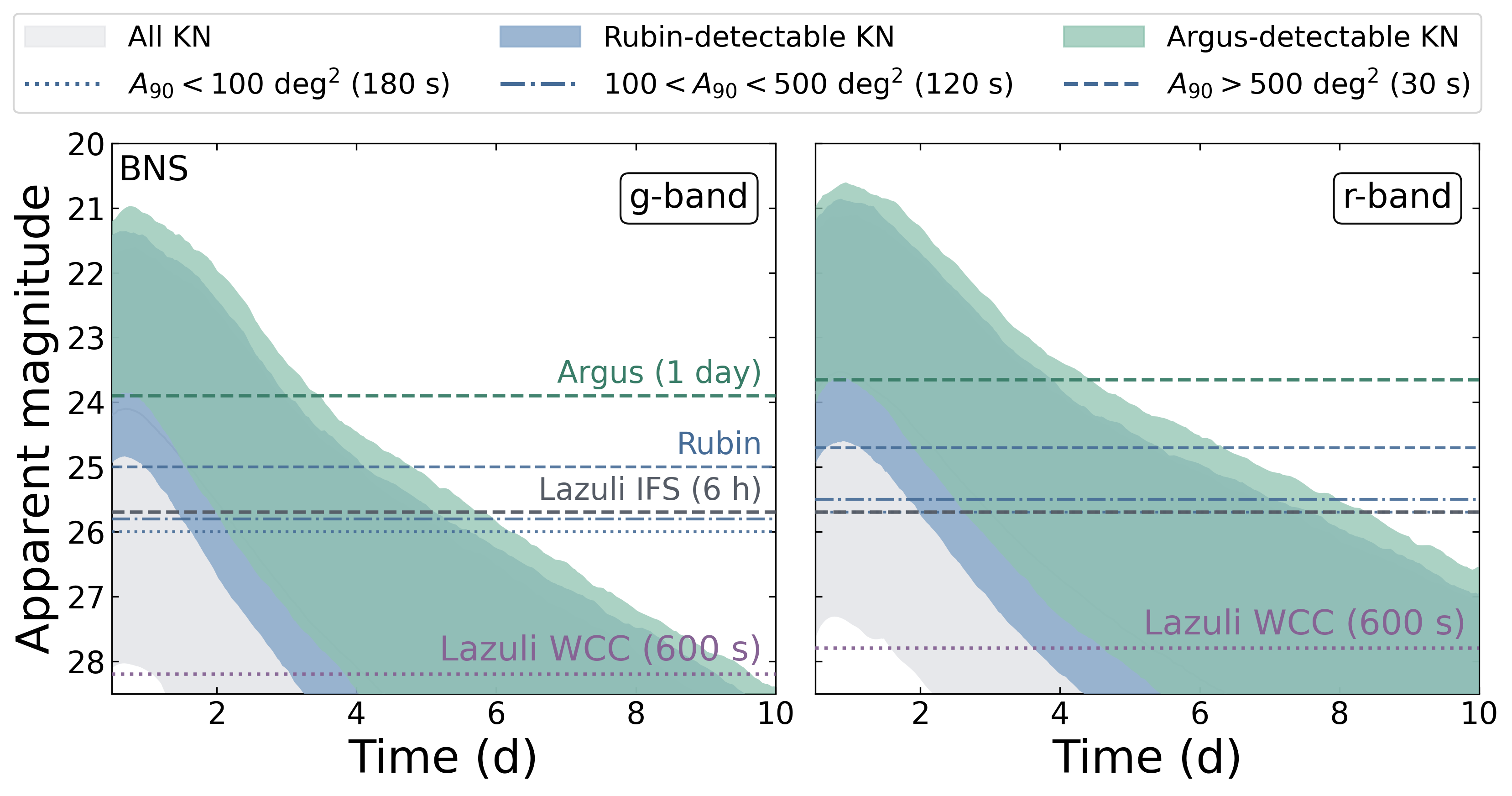}
\includegraphics[scale = 0.56]{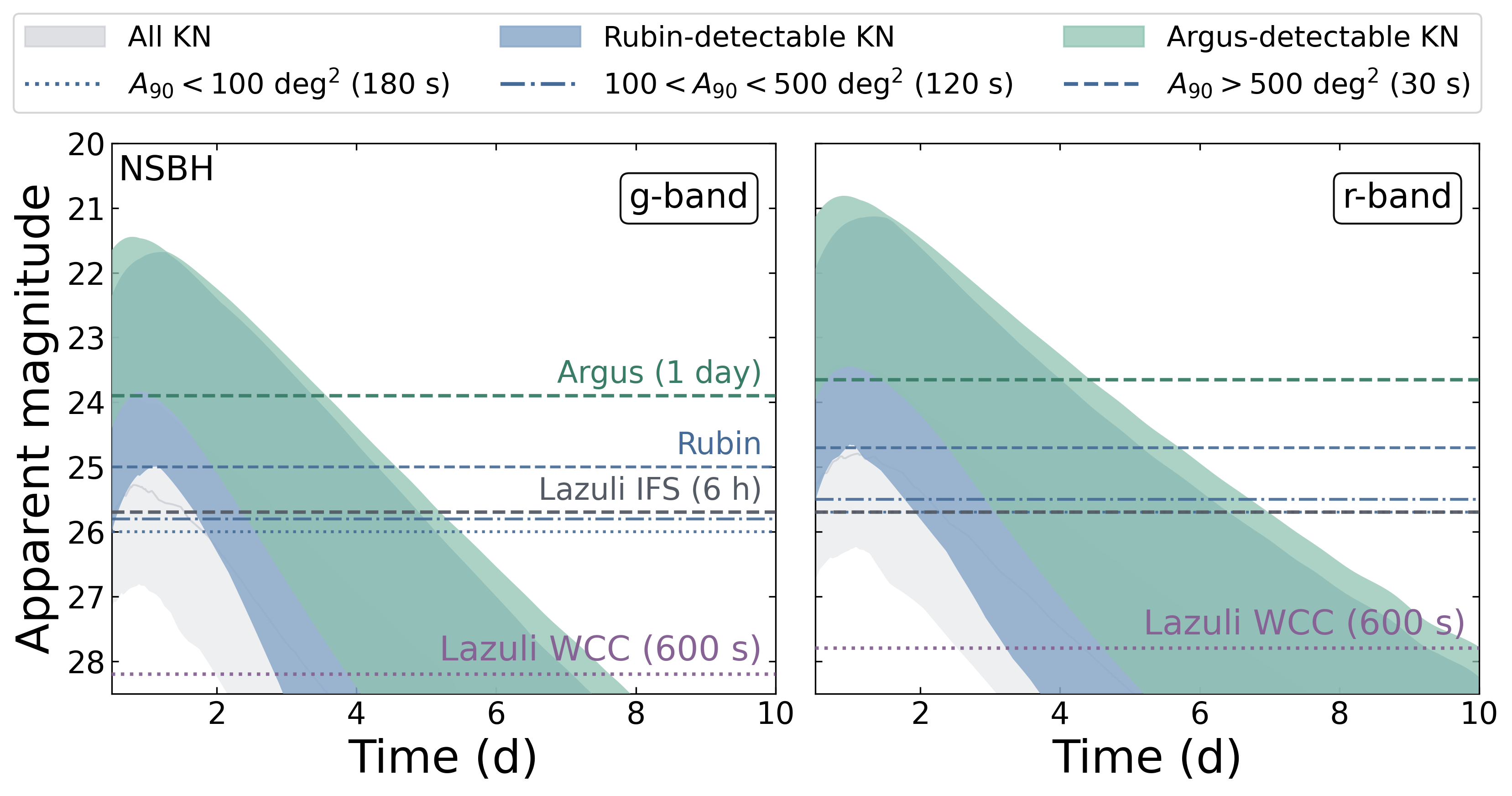}
\caption{Simulated BNS (top) and NSBH (bottom) kilonova lightcurves in g-band (left) and r-band (right). The grey shaded region shows the central 90th percentile of all simulated kilonova lightcurves, whereas the green and blue shaded region shows the 90th percentile of the kilonova events detectable by Argus and Rubin respectively. The green, black and violet dashed lines indicate the adopted limiting magnitudes for the Argus 1 day coadd, 6 hour Lazuli IFS exposure and 600 seconds Lazuli WCC exposure (assuming g' band with a HWK4123 low read noise detector). The three blue dashed, dot-dashed, and dotted lines show the limiting depths for Rubin adopted in the ToO strategy based on the source localization: Rubin 30, 120 and 180 seconds, respectively.}
\label{fig:lcbns_nsbh_r_g}
\end{figure*}

In Table \ref{tab:o5_rates_kn_detectability}, we show BNS and NSBH GW and kilonova detectabilities assuming LVK O5 sensitivity. Considering a BNS detection rate of $\sim 1.2 - 12.3$ yr$^{-1}$ and the kilonova detection efficiencies from the simulations, we estimate BNS kilonova detection rates for Argus and Rubin. We find a kilonova detection rate of $0.3 - 4.2$ yr$^{-1}$ with Argus and $0.6 - 6.6$ yr$^{-1}$ with Rubin for a total of $0.7 - 8.2$ yr$^{-1}$. For NSBH, considering a GW detection rate of $6.0 - 25$ yr$^{-1}$, we find a kilonova detection rate of $0.0 - 0.2$ yr$^{-1}$ with Argus and $0.2 - 0.7$ yr$^{-1}$ with Rubin, for a total  of $0.2 - 0.8$ yr$^{-1}$. Since NSBH mergers are usually more massive than BNS systems, they can be detected much farther in gravitational waves and hence their EM counterparts, even if present, might be too faint for the telescopes to detect, in addition to the low fraction of kilonova-producing NSBH.

For each simulated GW event, the source's sky position determines its observability by Argus and Rubin. Observable events are required to have an airmass $< 2.2$ during the first 24 hours following the event and to satisfy the adopted limiting magnitude cuts in the relevant bands. Figure \ref{fig:lcbns_nsbh_r_g} presents the resulting BNS and NSBH kilonova lightcurves. The grey region shows the 5-95\%-ile of the simulated O5 BNS and NSBH kilonova populations. The green (blue) region shows the 5-95th percentile of the subset of kilonovae detectable by Argus (Rubin) within the first 5 days of GW trigger in g-band (left) and r-band (right). 

As shown in Figure \ref{fig:lcbns_nsbh_r_g}, the 95 \%-ile of the Argus-detectable BNS kilonova lightcurves falls below the detection threshold for the 1 day coadd within $\sim 3.5$ days in $g$-band and within $\sim 4.2$ days in $r$-band. For BNS events with 90\% localization area $A_{90}< 100$ deg$^{2}$, the kilonova lightcurves are detectable up to 5.5 days in $g$-band up to 7.6 days in $r$-band with Rubin. {If $100 < A_{90} < 500$ deg$^{2}$, kilonova lightcurves are detectable within $\sim 5.2$ and $\sim 7.0$ days in $g$ and $r$-bands respectively.} For poorly localized events with 90\% sky area $>$ 500 deg$^{2}$, 95\% of the kilonova lightcurves are detectable by Rubin within the first $\sim 4.1$ days in $g$-band and within $\sim 5.4$ days in $i$-band with 30 s exposure. If we require reaching an SNR $>$ 5 in 6 hours ET, these objects are observable by Lazuli IFS up to $\sim 6-7$ days from the GW merger. We find that $\sim 28\%$ of the BNS kilonovae are within the Argus-detectable range and $\sim 48\%$ are detectable by Rubin, with an overlap of $\sim 18\%$.

NSBH kilonova lightcurves are shown in the bottom row of Figure \ref{fig:lcbns_nsbh_r_g}. Argus-detectable NSBH kilonovae can be observed  within the first $\sim 3.7$ days post merger. For events with sky localization $A_{90} < 100$ deg$^{2}$, Rubin can detect kilonova lightcurves for up to $\sim 5.0$ days in $g$-band and $\sim 6.4$ days in $r$-band. {For $100 < A_{90} < 500$ deg$^{2}$, $g$ and $r$-band observations are possible up to $\sim 5.0$ and $\sim 6.0$ days respectively.} For events with large localization areas ($A_{90} > 500 ~\rm deg^{2}$), detection is typically only possible within the first 4.2 (5.2) days in $g$ ($r$) band. {About $7\%$ of the simulated NSBH population can produce a kilonova.  Out of those, $\sim 10\%$} are within the Argus detectable range and $\sim 36\%$ are within the Rubin detectable range with an overlap of $\sim 9\%$ between the 2 telescopes.

\subsection{Kilonova spectroscopic observations}

\begin{figure}[!htpb]
    \centering
    \includegraphics[width = \linewidth]
    {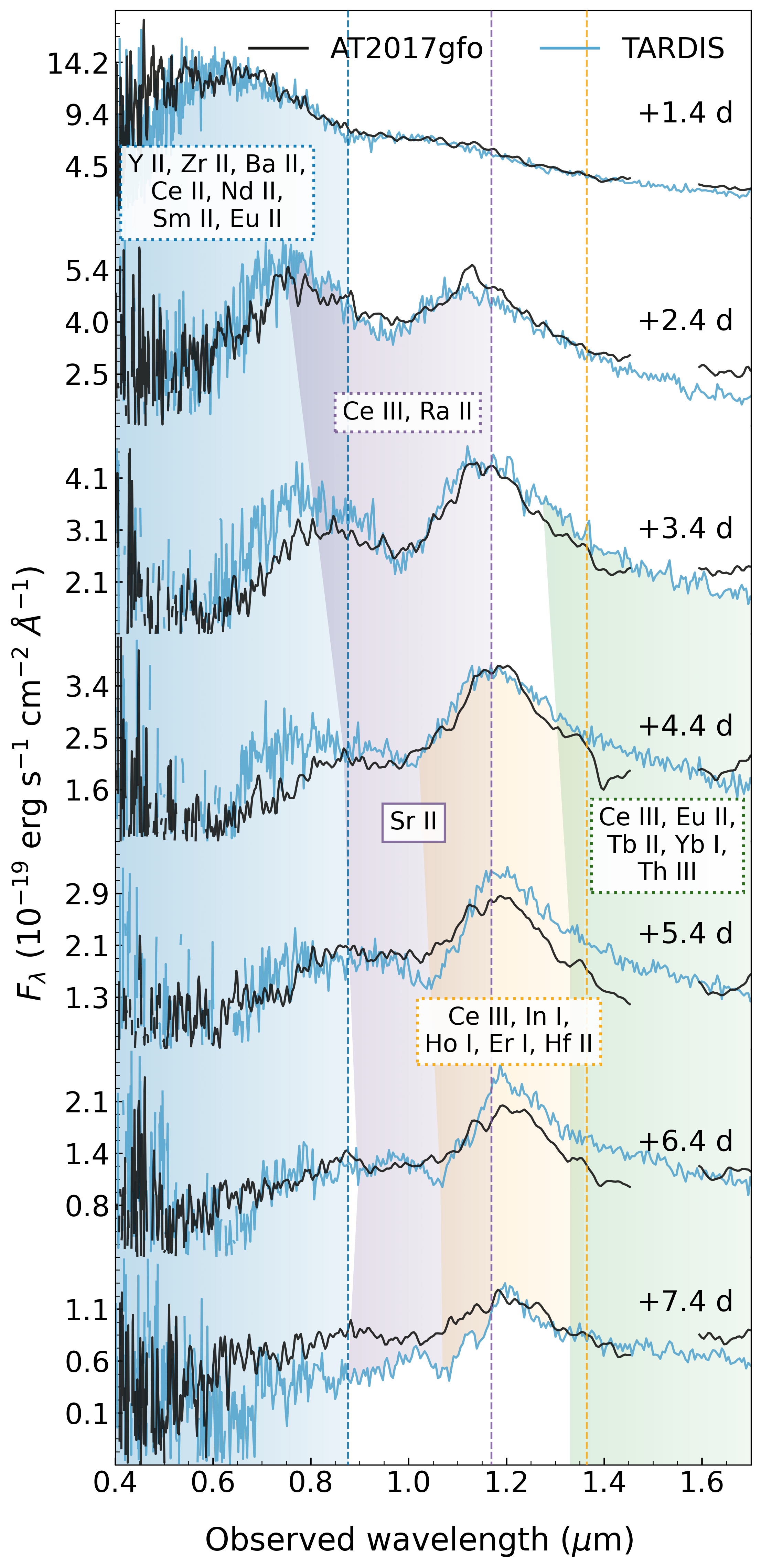}
    \caption{Lazuli/IFS spectra, simulated using \texttt{slicersim} (black), based on the AT2017gfo X-shooter data shifted to a luminosity distance of 521 Mpc. Telluric absorption lines (near 1.5 $\mu$m) are masked. Blue curves represent the best-fitting \texttt{TARDIS} models \citep{2022Gillanders,2024Gillanders}. Dashed lines mark the location of the proposed rest-frame features (redshifted to $D_{L}$ = 521 Mpc). The shaded regions blueward of these lines indicate the wavelength ranges over which the absorption components of possible P-Cygni profiles are expected, with the shading intensity increasing with blueshift. Ions with confirmed detections are shown in solid boxes, while the dotted boxes show the candidate ions whose identification remains uncertain.}
    \label{fig:tardis_spectra}
\end{figure}

We simulate the spectral evolution of a kilonova based on the X-shooter data of AT2017gfo \citep{Pian2017,Smartt2017} from 1.4 to 7.4 days after the merger (Figure \ref{fig:tardis_spectra}), shifted to a distance of 521 Mpc (which is the median Argus detectable kilonova distance from our simulations). Black curves represent simulated Lazuli/IFS spectra (generated using \texttt{slicersim}); to achieve SNR = 20 over the spectral range 0.7--1.7 $\mu$m, typical exposure times range from 0.8 to 6.0 hours. The blue curves in Figure \ref{fig:tardis_spectra} show the best-fitting \texttt{TARDIS} model (fit to the simulated Lazuli data) for each of the epochs \citep{2022Gillanders}. At 1.4 days, the best-fitting ejecta composition corresponds to $Y_e = 0.37$ with a photospheric velocity $v_{\rm min} = 0.28c$. From $2.4 - 7.4$ days, the best-fitting parameters correspond to a $Y_e = 0.29$ and $v_{\rm min}$ of 0.20c, 0.15c, 0.12c, 0.09c, 0.07c and 0.05c for 2.4, 3.4, 4.4, 5.4, 6.4 and 7.4 days respectively. Assuming a maximum ET of 6 hours, in the 7.4 day epoch Lazuli/IFS could reach a mean SNR of 14 for the X-shooter spectra and an SNR of 12 for the best-fit TARDIS model. The IFS sensitivity is lowest in the blue end, leading to relatively noisy spectra; however, features can clearly be identified by eye in the red part ($>$ 0.7 $\mu$m) of the spectrum at all epochs.

\begin{figure}[htpb!]
\centering
\includegraphics[scale = 0.44]{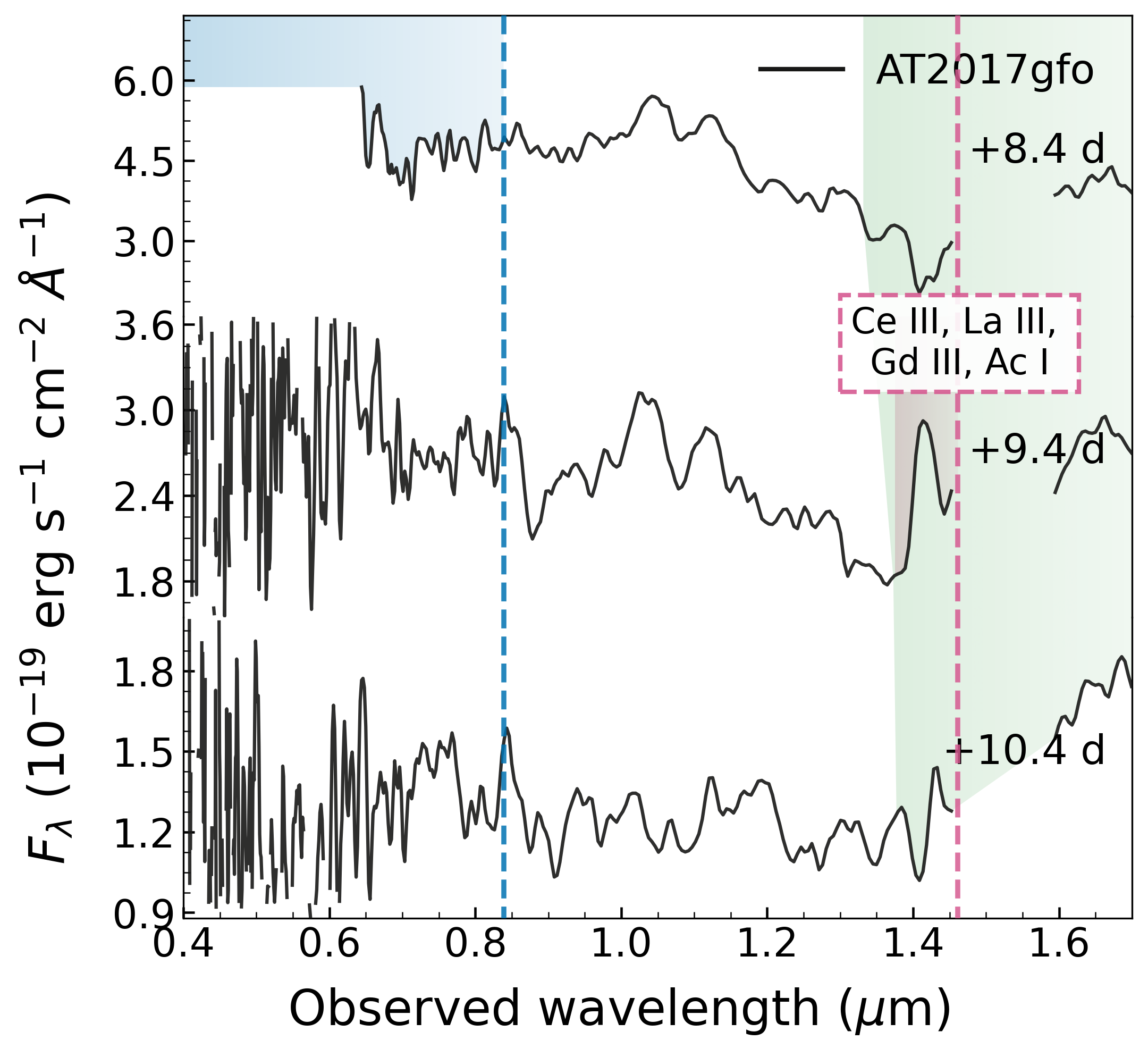}
\caption{Same as Figure \ref{fig:tardis_spectra}, but for the later (and fainter) epochs 8.4, 9.4 and 10.4 days after the merger; in this case we have shifted the event to 200 Mpc instead of 521 Mpc to achieve comparable SNR.}.
\label{fig:late_spectra_d200}
\end{figure}

\begin{figure*}[!htpb]
\centering
\includegraphics[scale = 0.45]{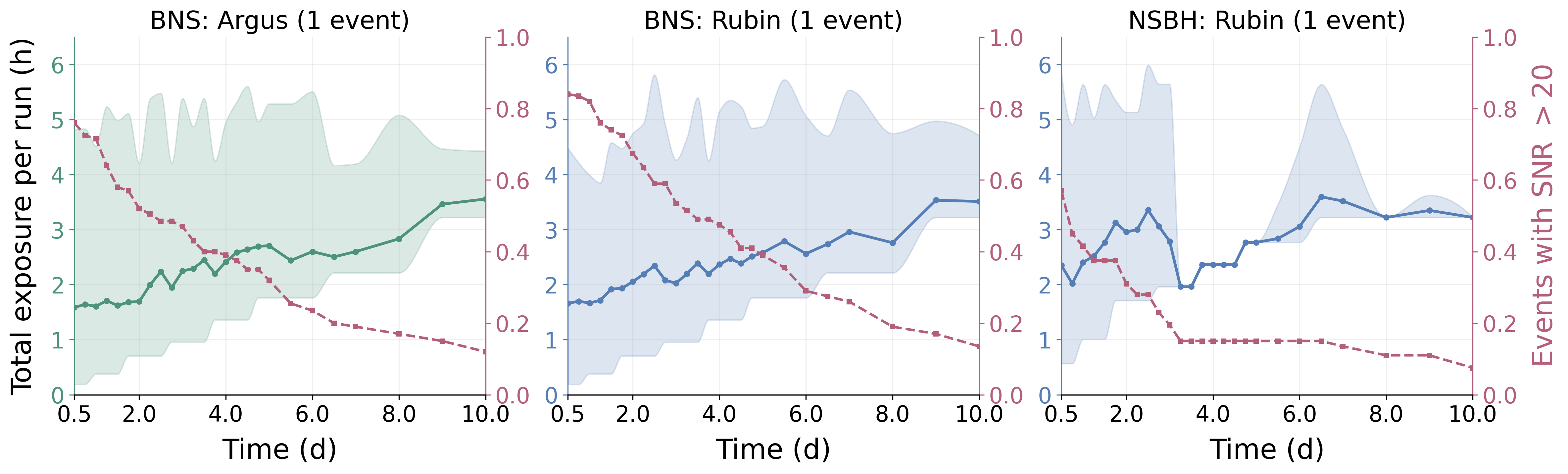}
\caption{Lazuli/IFS ET required at each post-merger epoch to follow up 1 Argus detectable BNS kilonovae, 2 Rubin detectable BNS kilonovae and 1 Rubin detectable NSBH kilonova respectively. The shaded region indicated the 5th-95th percentile whereas the green/blue lines represent the mean ET per epoch. These ET values are averaged across 50 random selections of Argus and Rubin detectable kilonova events. The red curve shows the fraction of events that passes the SNR cut of 20.}
\label{fig:etc_hist}
\end{figure*}

As discussed in \citet{2022Gillanders} and \citet{2024Gillanders}, it would be difficult to identify individual spectral lines for the observed features because the relative line strengths of the candidate ions are poorly constrained by existing atomic data. Moreover, incomplete atomic data for heavy r-process elements and lack of well-calibrated atomic data would make definitive line identification challenging for the observed kilonova spectral features. In addition to that, large expansion velocities associated with the kilonova ejecta render line-blending common in the observed spectra and affect accurate line identification. Thus, definitive line identifications do not depend on observability alone and as such we do not claim that Lazuli will be able to unambiguously identify features despite blending. We rather focus on the identification of features, even though their origin will need to be further analyzed with future improved atomic data and kilonova modeling.

Figure \ref{fig:tardis_spectra} illustrates that Lazuli/IFS spectra would enable the identification of broad spectral features as observed in AT2017gfo at much larger distances. The shaded regions blueward of the dashed lines show the wavelength region over which we expect to see the absorption features corresponding to possible P-Cygni profiles at a range of velocities. Ions with confirmed detections are shown in solid boxes whereas the candidates that might be contributing to the feature within the shaded region are listed in dotted boxes. The blue shaded region experiences significant lanthanide line blanketing \citep{2021kawaguchi,2022Ristic}and is dominated by lanthanides such as Ce, Nd, Sm and Eu \citep{2024Gillanders}. Sr\,\textsc{ii} has been identified as the ion contributing to the feature identified within the violet shaded region in the first few days ($\sim 7.4$ days) \citep{2019Watson, 2021domoto,2022Perego, 2022Gillanders}, with Ce\,\textsc{iii} and Ra\,\textsc{ii} being the alternative explanation for the same feature. Ce\,\textsc{iii} is listed as the most likely candidate for the yellow shaded region \citep{2022domoto,2024Gillanders} whereas the proposed candidates for the green shaded region include lanthanides such as Ce\,\textsc{iii}, Eu\,\textsc{ii}, Tb\,\textsc{ii}, Yb\,\textsc{i} and actinides such as Th\,\textsc{iii}. \citep{2022domoto,2024Gillanders}

Even though Sr\,\textsc{ii} may dominate the violet-shaded feature at early times, \citet{2024Gillanders} and \citet{2023Tarumi} support the possibility of He\,\textsc{i} becoming a dominant ion at later times (7.4 day onwards). To observe an AT2017gfo-like kilonova at a distance of 200 Mpc (see Figure \ref{fig:late_spectra_d200}), Lazuli would require an ET of 1.1, 2.4 and 5.6 hours to reach an SNR of 20 for epochs 8.4, 9.4 and 10.4 days respectively. The location of the proposed rest-frame features, but redshifted for a distance of 200 Mpc are marked in the figure. As described in \citet{2024Gillanders}, identifiable features lie near 0.8 and 1.5 $\mu$m at 8.4 days, while prominent peaks occur near 1.4 (identified by \citealt{Tanvir2017,2023Sneppen}) and 1.58 $\mu$m at 9.4 days. At 10.4 days, only the 1.58 $\mu$m feature remains identifiable in the simulated spectra at SNR $>$ 20. To observe these late time features in AT2017gfo-like kilonova at larger distances ($\ge$400 Mpc), a 6 hour ET would result in an average SNR of 16, 10 and 6 respectively for each epoch. At the median Rubin detectable kilonova distance of 600 Mpc, SNR $>$ 5 can only be achieved up to 8.4 days within 6 hours. 

The 0.5 day spectra is a hot blue featureless continuum, as described in \citet{Shappee2017}, so we do not discuss it further here.  

\begin{figure}[!htpb]
\centering
\includegraphics[scale = 0.115]{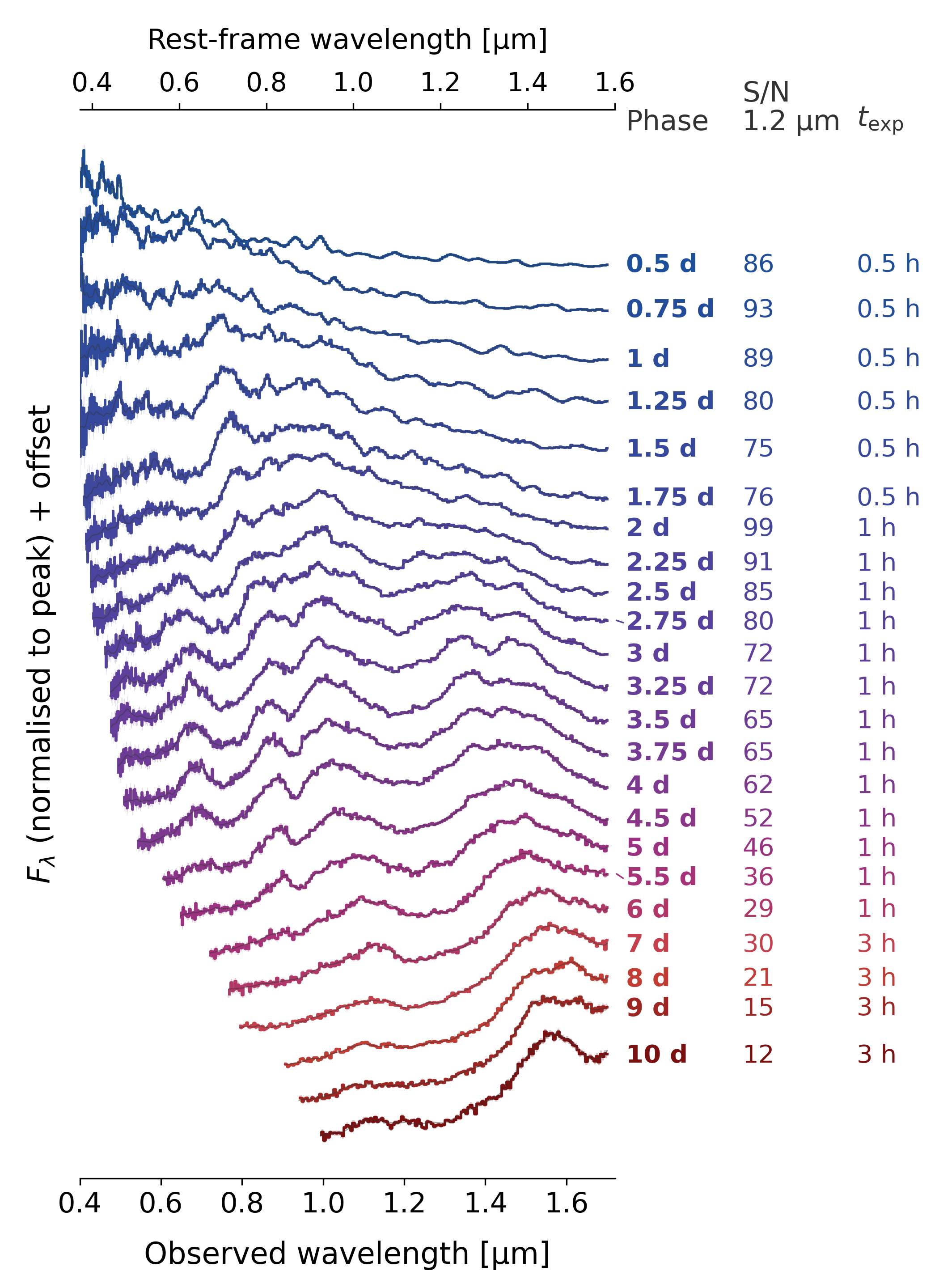}
\caption{{Lazuli/IFS simulations of a model kilonova spectra at 350 Mpc, simulated using \texttt{slicersim}. The selected kilonova has M$_{\rm dyn} = 0.005 \rm M_{\odot}$, M$_{\rm wind} = 0.050 \rm M_{\odot}$, v$_{\rm dyn}$ = 0.25c, v$_{\rm wind}$ = 0.10c at Y$_e$ = 0.30 and $\theta_v = 41.7\degree$. The ETs required to reach the respective SNRs for each epoch (at 1.2 $\mu$ m) are shown on the right side of the plot.}}
\label{fig:spectra_bnskn_grid}
\end{figure}

\subsection{Time estimates for Lazuli kilonova follow-up}
\label{kn_time_estimate}

\begin{figure*}[htpb!]
\centering
\includegraphics[scale = 0.47]{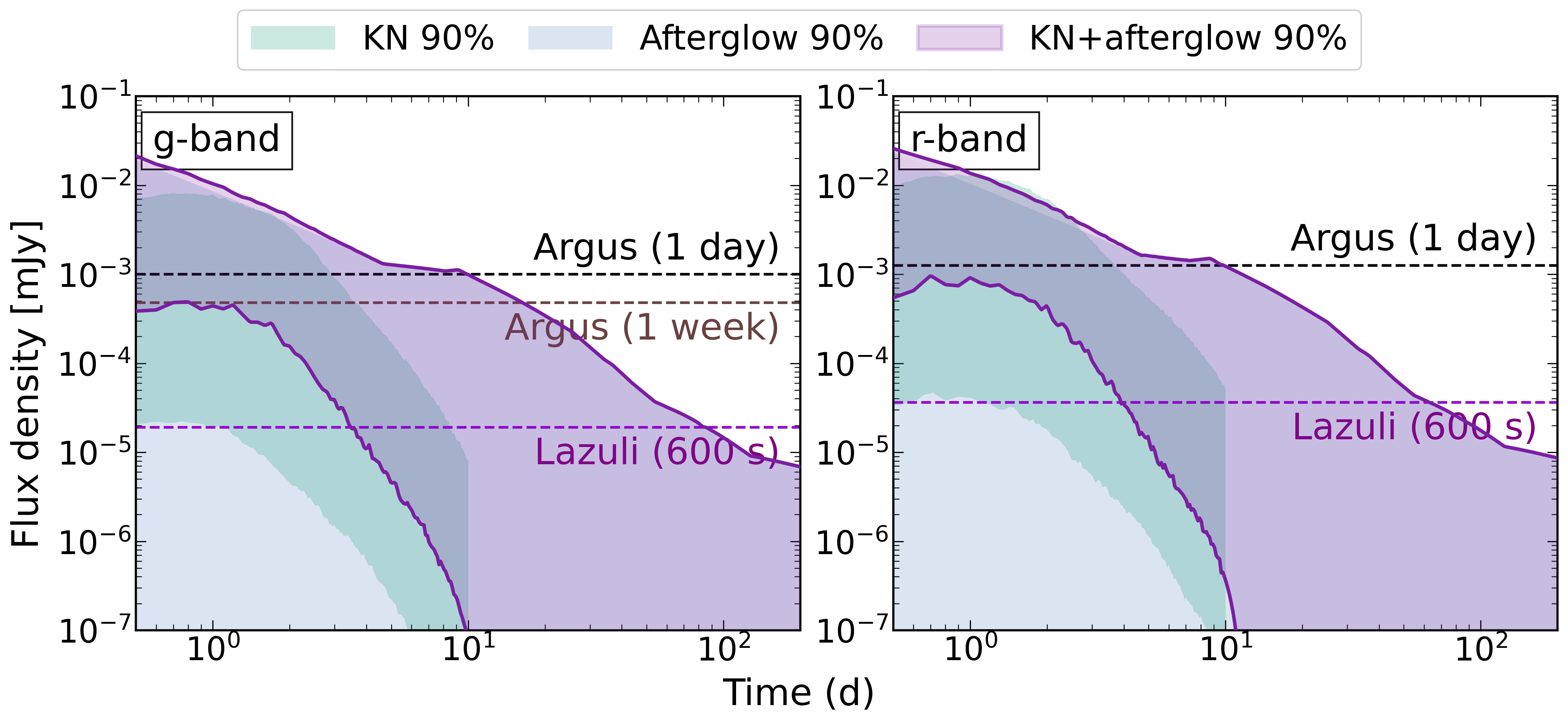}
\caption{Central 90th percentile of {all the Argus detectable} kilonova and afterglow lightcurves in the g-band (left) and r-band (right). From 0.5 to 10 days, the flux density is plotted as the sum of kilonova+afterglow components (purple region); beyond 10 days, only the afterglow component is considered. The 90th percentile of kilonova lightcurves are shaded in green for reference. The horizontal lines denote the adopted Argus and Lazuli WCC $5\sigma$ detection limits.}
\label{fig:kn_afg_combined_g_r}
\end{figure*}

A follow-up strategy with Lazuli should consist of two components: IFS spectroscopy of the kilonova and broadband imaging follow-up of the kilonova and afterglow (the WCC can cover the $u$'-band, which is blueward of the IFS spectral range). Figure \ref{fig:etc_hist} shows the Lazuli IFS ETs required to reach a mean SNR of 20 for BNS kilonovae detectable by Argus and by Rubin, from 0.5 to 10 days post merger. The SNR is averaged over 0.7 to 1.7 $\mu$m, i.e. excluding the lower sensitivity blue end of the IFS range. 

Each subplot shows the distribution of exposure time needed for Lazuli to take low SNR follow-up spectra, calculated from 50 random draws of detectable events. Each draw contains the median number of events expected per year during O5 (1 for Argus, 2 for Rubin for BNS and 0 for Argus, 1 for Rubin for NSBH; Table \ref{tab:o5_rates_kn_detectability}). 

The secondary axis (red curve) shows the number of events for which Lazuli can achieve SNR$>$5 in $<$6 hours of exposure time; these are the events where Lazuli could realistically obtain at least 1 spectrum. {For Argus detectable events, ETs per epoch varies between 0.2 - 1.8 hours for the first few days to 3.2 - 5.0 hours for the later epochs. For Rubin detectable BNS kilonovae, the total ET per epoch for detecting 2 events yr$^{-1}$ is $\sim 0.2-3.8$ hours within the first 24 hours and 6.5-11 hours at later epochs. For Rubin detectable NSBH kilonovae, the average ET ranges from 1.5 hour in the first 24 hours to 3.2 hours per event from 6 days from merger.}

To highlight the high cadence observations that Lazuli can obtain from space, in Figure \ref{fig:spectra_bnskn_grid} we show a spectral series for a randomly selected BNS kilonova from our simulations at 350 Mpc. The spectra are normalized and offset for clarity. We show the spectra from 0.5 to 10 days, with a cadence of 6 hours between exposures within the first 4 days, 12 hours from 4 - 6 days, and 1 day cadence thereafter. Significant evolution in the SED as well as spectral features can clearly be distinguished in the IFS spectra, including P-cygni profiles and emission features. High cadence, high precision spectrophotometric time-series will allow self-consistent modeling to constrain the dynamics and physics of kilonovae.

The IFS observing sequence for this particular event would need $\sim$28 hours of Lazuli time. Considering a median of 3 Argus+Rubin detectable BNS kilonovae per year in O5 (see \ref{tab:o5_rates_kn_detectability}), this would call for a total ET requirement of $\approx 100$ hours per year for Lazuli for spectroscopic follow-up (but depending sensitively on the distance distribution). At 350 Mpc, a 3 hour spectrum at 10 days post-merger would still result in an SNR of 12. Other observatories such as JWST will be able to continue monitoring out to later times. 

In our NSBH kilonova simulations, most prominent features appear at redder wavelengths (1.4 $\ge \mu$m), largely outside the Lazuli/IFS wavelength coverage, hence we do not show the spectra here. Although several kilonova predictions agree that NSBH kilonovae tend to be redder \citep{2015Kasen,2017CFern}, we haven't observed a confirmed NSBH kilonova and we cannot exclude a blue ejecta component, particularly during the early epochs when viewed close to the polar axis \citep{2021Kyutoku,2023Curtis}. Lazuli follow-up of NSBH kilonovae, could therefore help understand the ejecta composition and test these predictions.

\begin{table*}[htpb!]
\centering
\begin{tabular}{ccccc}
\hline
\textbf{Telescope} &
\textbf{Threshold} &
\textbf{Frequency} &
\textbf{Afterglow detectability} & \textbf{Afterglow detection rate} \\
& (mJy) & (Hz) & (\%) & ($\rm yr^{-1}$) \\
\hline

Chandra (50 ks)
    & $4.40\times10^{-7}$
    & $2.42\times10^{17}$
    & 5.85 & 0.04 - 0.43\\
Argus $g$-band (1 d)
    & $1.00\times10^{-3}$
    & $6.21\times10^{14}$
    & 3.05 & 0.02 - 0.22\\
Argus $g$-band (6 mo)
    & $9.12\times10^{-5}$
    & $6.21\times10^{14}$
    & 5.16 & 0.04 - 0.38\\
Argus $r$-band (1 d)
    & $1.26\times10^{-3}$
    & $4.76\times10^{14}$
    & 3.03 & 0.02 - 0.22\\
Lazuli $i$-band (40 min)
    & $3.31\times10^{-5}$
    & $3.95\times10^{14}$
    & 7.28 & 0.05 - 0.53\\
Lazuli broadband (40 min)
    & $1.66\times10^{-5}$
    & $5.00\times10^{14}$
    & 8.32 & 0.06 - 0.60\\
Rubin $g$-band (30 s)
    & $4.79\times10^{-4}$
    & $6.21\times10^{14}$
    & 3.55 & 0.03 - 0.26\\
Rubin $i$-band (30 s)
    & $1.10\times10^{-3}$
    & $4.00\times10^{14}$
    & 3.19 & 0.02 - 0.23\\
Roman $R$-band (67 s)
    & $1.20\times10^{-4}$
    & $4.87\times10^{14}$
    & 5.06 & 0.04 - 0.37\\
Roman $J$-band (67 s)
    & $2.29\times10^{-4}$
    & $2.32\times10^{14}$
    & 4.96 & 0.04 - 0.36\\
DSA (1 hr)
    & $1.80\times10^{-3}$
    & $1.40\times10^{9}$
    & 27.40 & 0.18 - 1.96\\
DSA (6 hr)
    & $7.30\times10^{-4}$
    & $1.40\times10^{9}$
    & 34.76 & 0.23 - 2.48\\
VLA (2 hr)
    & $7.00\times10^{-3}$
    & $6.00\times10^{9}$
    & 12.93 & 0.12 - 1.27\\
\hline
\end{tabular}
\caption{Afterglow detectability for different telescopes out of the BNS events that launch a jet. The detection rate uncertainties are derived from the 90\% credible interval of the BNS merger rate (see Eq. \ref{eq:2}), rescaled to account for no BNS detection in O4.}
\label{tab:afterglow_detectability}
\end{table*} 

\subsection{Afterglow detectable sources}
\label{afg_sources}

Figure \ref{fig:kn_afg_combined_g_r} shows the combined kilonova and afterglow lightcurves in the Argus $g$ (left panel) and $r$-band respectively (right panel), for a sample of BNS events that satisfy both the kilonova and afterglow detection criteria (independently) in the respective bands. For each of the selected BNS events that launch a jet, the kilonova lightcurves are computed from 0.5 to 10 days using \texttt{NMMA}, while the afterglow lightcurves are computed using run 7 parameters of \citet{2026Kaur}. The 5-95 percentile of the afterglow lightcurves remain detectable in Lazuli $g$-band up to $\sim 35$ days after the BNS merger, compared to $\sim 25$ days in Lazuli $r$-band. $\sim70$\% of the Lazuli detectable afterglows have a kilonova detectable by either Argus or Rubin, which supports a pointed stare corresponding to the sky location reported for the kilonova.

Of the simulated BNS mergers, 58\% successfully launch a jet; Table \ref{tab:afterglow_detectability} gives the fraction of these that produce a detectable afterglow for each telescope. Across all wavelengths, radio telescopes achieve the highest detection fractions, with DSA detecting 35\% (at an ET of 6 hours) and 27\% (at an ET of 1 hour) and VLA detecting (at an ET of 2 hours) 13\% of the jet launching population, respectively. This could result in an afterglow detection rate of up to $\sim 2-3$ per year for DSA. At optical and near-infrared wavelengths, Roman outperforms Rubin in both bands, while Lazuli achieves detection rates of $\sim 7\%$ and $8\%$ in i-band and broadband (BB) {all of them detectable by DSA}, making it a competitive instrument for afterglow follow up alongside DSA. At least for a fraction of events, measuring a photon spectral index from the multi-wavelength observations from Schmidt observatories will thus be possible.

\begin{figure}[htpb!]
\centering
\includegraphics[scale = 0.5]{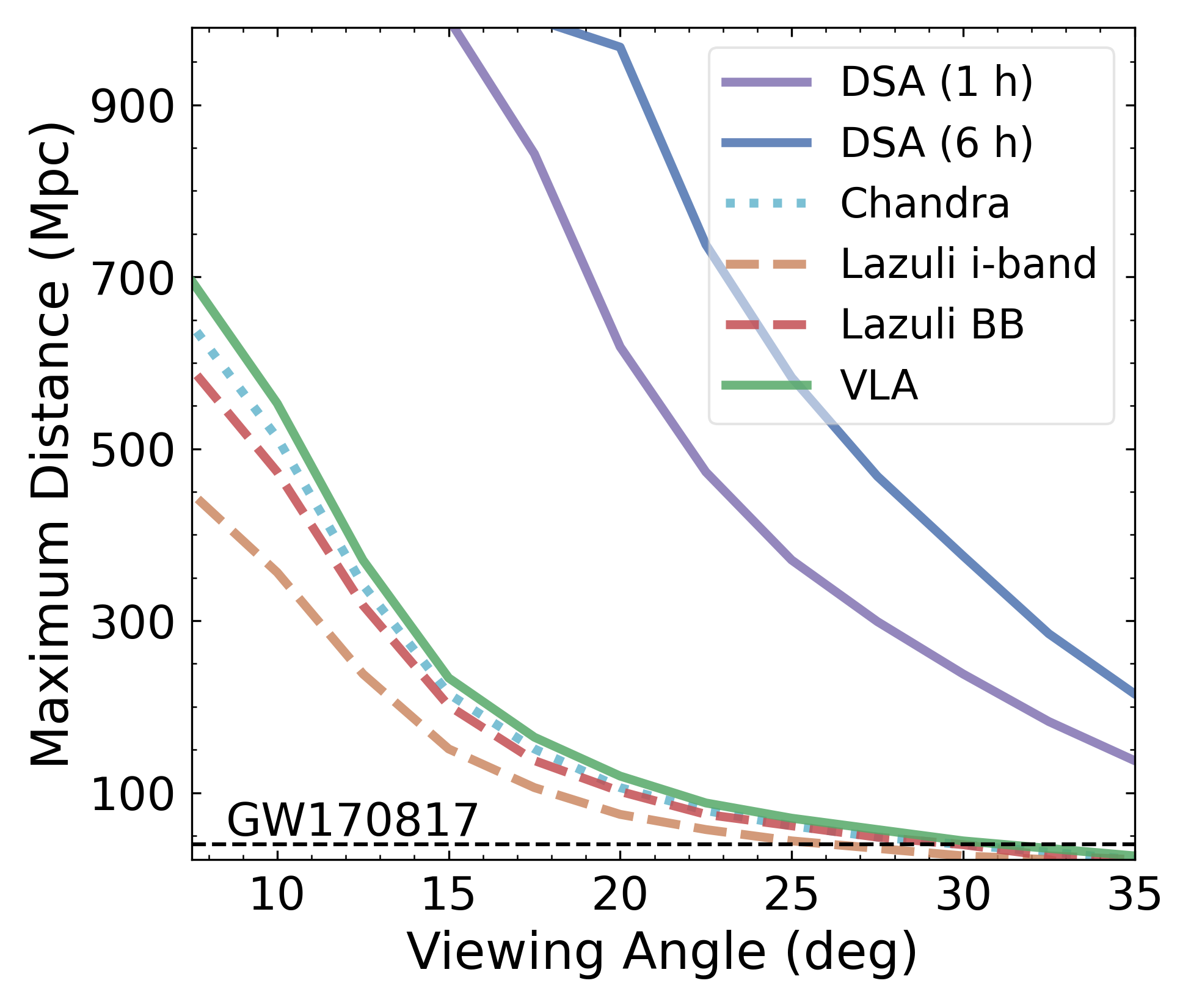}
\caption{Maximum detectable distance of GW170817-like jet afterglows as a function of viewing angle for different facilities.}
\label{fig:theta_v_afg}
\end{figure}

\begin{figure*}[htpb!]
\centering
\includegraphics[scale = 0.48]{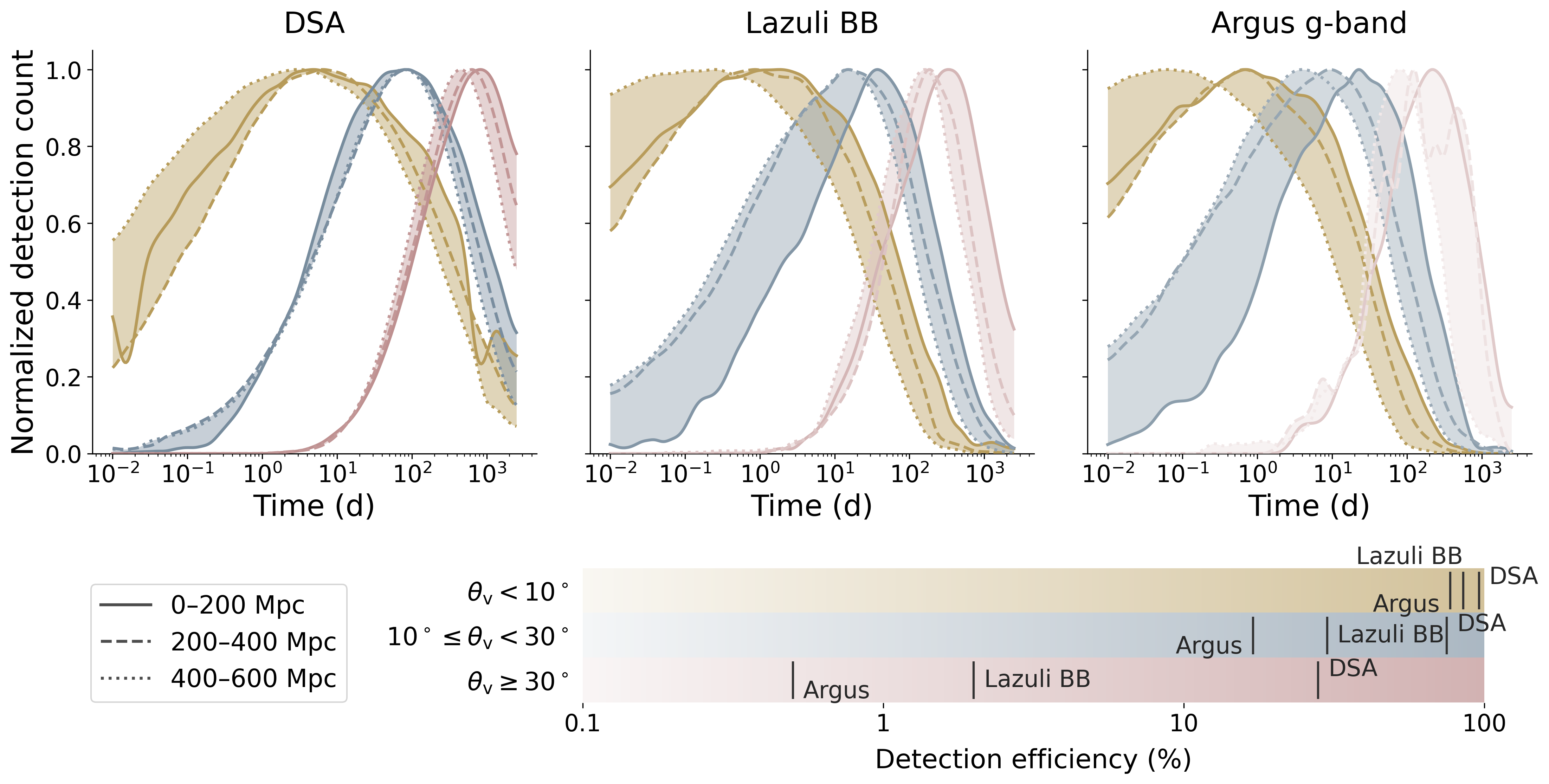}
\caption{{Fraction of jet afterglows that are detectable above the DSA (6h), Lazuli BB (40 min) and Argus g-band (1 day co-add for the first 7 days, 1 week co-add from 7 days to 6 months, and 6 month co-add thereafter) detection limit, up to 1000 days since the BNS merger. Afterglows are binned into 3 categories based on how far the event is located, and they are further classified based on their viewing angle. The shading intensity indicates the afterglow detection efficiency within each viewing angle and distance bin. Markers on each colorbar indicate the mean detection efficiency of each telescope within the corresponding viewing angle range.}}
\label{fig:det_count_time_afg}
\end{figure*}

Figure \ref{fig:theta_v_afg} shows how the maximum detectable distance changes with viewing angle for different observing facilities and a GW170817-like afterglow \citep{Ryan2024}. Since afterglows are intrinsically brighter when viewed closer to the jet axis, the detectability decreases rapidly with increasing $\theta_{v}$. Lazuli, Chandra, and VLA can detect a GW170817-like afterglow viewed 10 deg off-axis out to distances of $\sim 450$, 470, and 550 Mpc, respectively, while their reach declines rapidly for off-axis systems. For DSA, events viewed 10$\degree$  off-axis remain detectable beyond 1 Gpc, whereas events at intermediate viewing angles, such as 30$\degree$, can be viewed up to a distance of 250 Mpc with a canonical 1 hour integration time and up to 400 Mpc with a 6 hour DSA exposure. Although we do not present an exhaustive list of observatories here, it is also clear, in comparison with the numbers of \citet{2026Kaur} that the DSA will significantly expand the discovery space for neutron star merger jet afterglows (see also \citealt{hallinan2019dsa2000radiosurvey}). 

65\% of the afterglows detectable by DSA has a kilonova detection in either Rubin or Argus. Afterglows with no KN detection tend to be farther with a median distance of 1 Gpc, but have favorable jet orientations (median viewing angle of 30$\degree$). In contrast, the afterglow population with a detectable kilonova are at a median distance of $\sim 500$ Mpc, with a median viewing angle of 40$\degree$. Regardless, our simulations shows that it would be worthwhile for DSA to follow-up such events as at a distance range of 800-1100 Mpc, 90\%, 66\%, 50\%, 38\%, 28\% and 20\% of afterglows are detectable for the viewing angle bin 0-10, 10-20, 20-25, 25-30, 30-35 and 35-40 $\degree$ respectively.

Note that the afterglow delectabilities reported do not account for the sky observability. On the other hand, most GW events are observed off-axis, and their afterglows tend to be slow and peak/are observable at late times (e.g. a GW170817-like afterglow at $\theta_v=10\degree$ is detectable for over 100 days by DSA). The afterglow is therefore significantly less affected than the kilonova by observability constraints.

\begin{figure*}[!t]
\centering
\includegraphics[scale = 0.52]{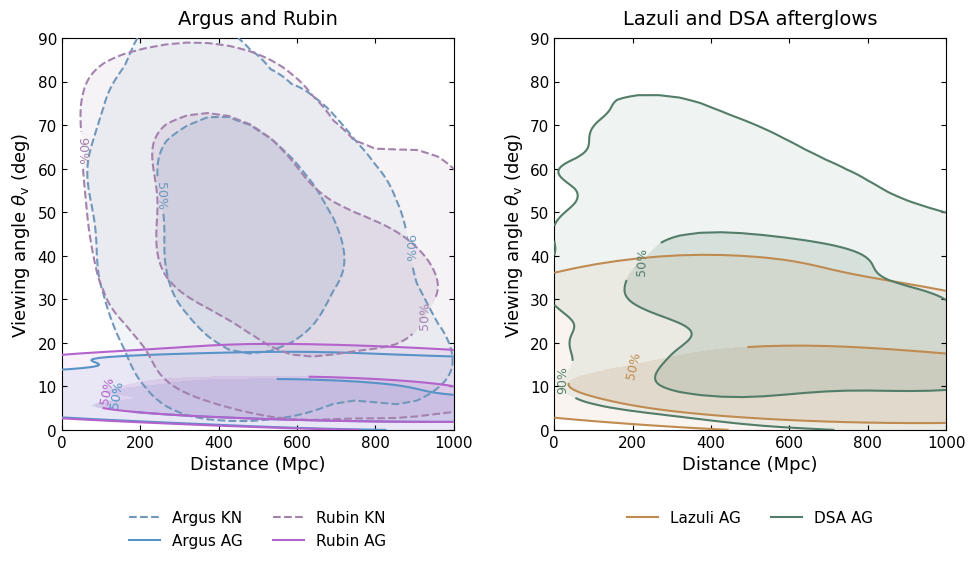}
\caption{{Distribution of simulated O5 BNS kilonovae and afterglow population in the viewing angle-distance space. In the left panel, the 50 and 90\% region of the blue and violet contour shows the kilonovae (dashed curve) and afterglows (solid curve) detectable by Argus and Rubin respectively, while the solid orange and green contours in the right panel represents the 50 and 90\% region of the afterglows detected by Lazuli and DSA respectively.}}
\label{fig:thetav_dis_contour_afg_kn1}
\end{figure*}

\subsection{Time estimates for afterglow follow-up}

For afterglow observations, DSA has the highest probability to detect a counterpart, and as shown in Figure \ref{fig:det_count_time_afg}, the observability peaks $\sim 5$ days from merger for face-on systems ($\theta_{v} < 10\degree$), at $\sim 80$ days for events at intermediate viewing angles ($10 < \theta_{v} < 20$) and $\sim 800$ days from the merger for more off-axis mergers ($\theta_v > 30\degree$). 

For Lazuli/WCC broadband observations, the detectability for face-on systems peaks within first $\sim 3$ days after the merger, whereas for intermediate $\theta_{v}$, the maximum number of detections is reached $\sim 15 - 40$ days after merger (modulo the distances at which the event occurred). For even more off-axis systems, events will have maximum detectability between $\sim$ 200 and 300 days from merger. If we consider Argus g-band detectability, face-on systems have the highest detectability within 0.7 hours to 1 day from the merger, whereas at intermediate $\theta_{v}$, systems can have peak detectability in Argus within $\sim 15$ days from merger. Off-axis systems that are bright enough will have maximum detectability $\sim 270-450$ days from the merger. 

On axis events are detected by all 3 telescopes with $>75\%$ efficiency. For events at intermediate viewing angles, detection efficiencies are 75\%, 30\% and 17\% for DSA, Lazuli and Argus respectively. For events with $\theta_v > 30\degree$, DSA detects 17\% of the events, Lazuli detects $\sim 2\%$ whereas Argus detects only 0.5\% of the off-axis events.

{Figure \ref{fig:thetav_dis_contour_afg_kn1} shows the parameter space that would allow for kilonova, afterglow or combined kilonova and afterglow detection. We show the 50 and 90\% regions of the kilonovae detectable by Argus and Rubin by dashed blue and violet contours (left panel), whereas the solid contours enclose the afterglows detectable by Argus (blue; left panel) and Rubin (violet; left panel), Lazuli (orange; right panel) and DSA (green; right panel). The KN regions are shaped by various effects: the GW selection effects, viewing angle dependence, volumetric effects, and the magnitude limit of the surveys. Lower viewing angle systems are easier to detect out to larger distances due to both the GW signal being louder, and the KN being brighter (within our assumed models). As expected, Rubin extends out to larger distances than Argus.

The 90\% Argus/Rubin detectable afterglow overlaps with the kilonova detectable region only for $\theta_v < 20\degree$; the same region is shared by 50\% of the Lazuli detectable afterglows. Lazuli's 90\% afterglow contour extends to $\theta_v$ of $<40\degree$. DSA reaches the furthest off-axis mergers: $\theta_v$ up to $75\degree$ within 300 Mpc, $65\degree$ at the median Rubin kilonova distance of 600 Mpc, and 50 degree at 1 Gpc.}

The detectability with Lazuli is highest in the first few days as optical on-axis afterglows are generally brightest at early times (see \ref{fig:det_count_time_afg}). Lazuli's kilonova epochs at 0.5-3 days already cover the face-on afterglow peak. So for face-on events, Lazuli could follow-up at high cadence in the first $\sim$ 4-5 days, and then adopt a sparser (roughly logarithmic) cadence beyond that (e.g. at 10 30, 100 and 200 days). For intermediate viewing angles, Lazuli's afterglow detections peak $\sim 15-40$ days and hence imaging for such events could focus around the first 10-50 days at a high cadence followed by a sparser cadence beyond. Later epochs, such as 100, 200 and 300 days, can be given priority for events with $\theta_v > 30\degree$ that are bright enough. Beyond $\theta_v = 40\degree$, the afterglow would be accessible only in the radio band, so we recommend coordinated DSA observations for such events. DSA reaches its maximum detectability for off-axis mergers at a few hundred days, as more off-axis jets become visible. Thus dedicated observations could be taken around 1, 10, 100, 300, 500, 800 and 1000 days from merger.
 
\section{Conclusions}\label{sec:conclusions}

In this work, we use BNS and NSBH mergers simulated at LVK O5 sensitivity to assess the detectability of GW-EM counterparts with the Schmidt Observatory System and complementary multi-wavelength facilities. We model the associated kilonova and relativistic afterglow emission and evaluate their detectability with Argus, Rubin, Lazuli and DSA. For the detectable kilonova population, we estimate the Lazuli exposure times required for multi-epoch spectroscopic follow up, using spectral models to assess the observability of key spectral features. We combine these results to develop a coordinated observing strategy extending from rapid kilonova follow-up to long term afterglow monitoring. Our main findings are as follows:

\begin{itemize}
    \item Approximately 28\% and 48\% of the simulated BNS kilonova population fall within the Argus and Rubin detectable samples respectively, with an overlap of $\sim 18\%$. For NSBH kilonovae, the corresponding fractions are $\sim 10\%$, $\sim 36\%$ and $\sim 9\%$.
    
    \item Most Argus detectable kilonovae remain above the 1 night co-add limits for approximately 3.5 days in the $g$-band and 4.2 days in $r$-band. Rubin extends this window to approximately 5 days depending on the localization area, filter and ET. We estimate that the combination of Argus and Rubin will discover $\sim 3-4$ GW-kilonovae per year (median of our simulations).
    
    \item For an AT2017gfo-like kilonova, the spectrum evolves from relatively featureless continuum at 0.5 days to a redder spectrum containing the prominent Sr II near-infrared triplet and possible late time forbidden transitions covered by Lazuli's IFS (0.4 - 1.7$\mu$m). At the median Argus kilonova detectable distance of 521 Mpc, Lazuli IFS reaches a target SNR of 20 in 0.8 hours at 1.4 days to 5.6 hours at 6.4 days, while achieving an SNR of 20. At 7.4 days, Lazuli could reach an SNR of 14 assuming a maximum exposure time of 6 hours.

    \item The rapidly evolving kilonova motivates dense Lazuli spectroscopic sampling at 0.5, 1, 2, 3, 4, 5 days, followed by observations at 7 and 10 days if the source is bright enough. Considering a median Argus and Rubin detection rate per year during O5, this would result in an average annual time allocation of $\sim28$ hours for each BNS kilonova, for a median of 3 kilonovae per year.
    
    \item DSA has the highest afterglow detectability among the facilities considered, detecting $\sim 27\%$ of BNS events that successfully launch a jet at 1 hour ET and $35\%$ at 6 hour ET (both for 3$\sigma$ limit), compared to $13\%$ for VLA and $7-8\%$ for Lazuli. For GRB170817A-like jets, face-on events are detectable by DSA beyond 1 Gpc, while an event viewed at 30$\degree$ is detectable up to a distance of $\sim 250$ Mpc with a 1 hour exposure and up to $\sim 400$ Mpc with a 6 hour exposure.

    \item Cadence for afterglow observations depends on $\theta_v$. Lazuli detectability is greatest in the first few days for favorably oriented events, when optical afterglows are brightest. DSA detectability peaks at $\sim 5$ days for on-axis events, $\sim 80$ days for intermediate viewing angles and $\sim 800$ days for $\theta_v > 30\degree$, as decelerating off-axis jets become visible at radio wavelengths.

    \item Lazuli's kilonova epochs at 0.5-3 days cover on-axis afterglows, with additional epochs closer to $\sim 30$ and $\sim 250$ days required for intermediate and slightly off-axis events. Beyond $\theta > 40\degree$ afterglows are only visible in the radio, so DSA could coordinate observations as more off-axis afterglows becomes visible between $300-1000$ days.

\end{itemize}

\nopagebreak
\section*{Acknowledgements}
AP and KK thank James Gillanders, Ravjit Kaur, Raffaella Margutti, Brendan O'Connor, David J. Sand and Thomas J. Maccarone for useful discussion. We also thank Mattia Bulla and James Gillanders for making their simulated kilonovae spectra available. The Lazuli Space Observatory is part of the Eric \& Wendy Schmidt Observatory System, an initiative of Schmidt Sciences.

\bibliography{references}
\bibliographystyle{aasjournalv7}

\end{document}